\documentclass[journal]{IEEEtran}
\usepackage{algorithmic}
\usepackage{array}
\usepackage[caption=false,font=normalsize,labelfont=sf,textfont=sf]{subfig}
\usepackage{textcomp}
\usepackage{stfloats}
\usepackage{url}
\usepackage{verbatim}
\usepackage{graphicx}
\def\BibTeX{{\rm B\kern-.05em{\sc i\kern-.025em b}\kern-.08em
    T\kern-.1667em\lower.7ex\hbox{E}\kern-.125emX}}
\usepackage{balance}
\usepackage{multicol}
\usepackage{multirow}
\usepackage{epsfig}
\usepackage{epstopdf}
\usepackage{latexsym}
\usepackage{amsfonts}
\usepackage{rawfonts}
\usepackage[T1]{fontenc}
\usepackage{calc}
\usepackage{url}
\usepackage{enumerate}
\usepackage{xcolor}
\usepackage{color}
\usepackage[tbtags]{amsmath}
\usepackage{amssymb}
\usepackage{upref}
\usepackage{times}
\usepackage{amsmath}
\usepackage{cite}
\usepackage{accents}
\usepackage{tikz}
\usepackage{url}
\usetikzlibrary{automata}

\newcommand{\bi}{\begin{itemize}}

\newcommand{\ei}{\end{itemize}}

\newcommand{\be}{\begin{enumerate}}

\newcommand{\ee}{\end{enumerate}}

\newcommand{\bd}{\begin{description}}

\newcommand{\ed}{\end{description}}

\newcommand{\bc}{\begin{center}}

\newcommand{\ec}{\end{center}}

\newcommand{\bt}{\begin{tabbing}}

\newcommand{\et}{\end{tabbing}}

\newcommand{\bfig}{\begin{figure}}

\newcommand{\efig}{\end{figure}}

\newcommand{\beq}{\begin{equation}}

\newcommand{\beqarr}{\begin{eqnarray}}

\newcommand{\beqarrn}{\begin{eqnarray*}}

\newcommand{\eeq}{\end{equation}}

\newcommand{\eeqarr}{\end{eqnarray}}

\newcommand{\eeqarrn}{\end{eqnarray*}}

\newcommand{\bflr}{\begin{flushright}\vspace{-0.2in}}

\newcommand{\eflr}{\end{flushright}}

\newcommand{\bsub}{\begin{subequations}}

\newcommand{\esub}{\end{subequations}}

\newcommand{\barr}{\begin{array}}

\newcommand{\earr}{\end{array}}

\begin{document}
\title{Jamming Intrusions in Extreme Bandwidth Communication: A Comprehensive Overview}
\author{Richa Priyadarshani,~\IEEEmembership{ Member,~IEEE}, Ki-Hong Park, \IEEEmembership{ Member,~IEEE}, Yalcin Ata, \IEEEmembership{ Member,~IEEE}, and  Mohamed-Slim Alouini,~\IEEEmembership{Fellow,~IEEE} 
\thanks{xx}
}


\maketitle

\begin{abstract}
As the evolution of wireless communication progresses towards 6G networks, extreme bandwidth communication (EBC) emerges as a key enabler to meet the ambitious key performance indicator set for this next-generation technology. 6G aims for peak data rates of 1 Tb/s, peak spectral efficiency of 60 b/s/Hz, maximum bandwidth of 100 GHz, and mobility support up to 1000 km/h, while maintaining a high level of security. The capability of 6G to manage enormous data volumes introduces heightened security vulnerabilities, such as jamming attacks, highlighting the critical need for in-depth research into jamming in EBC. Understanding these attacks is vital for developing robust countermeasures, ensuring 6G networks can maintain their integrity and reliability amidst these advanced threats. Recognizing the paramount importance of security in 6G applications, this survey paper explores prevalent jamming attacks and the corresponding countermeasures in EBC technologies such as millimeter wave, terahertz, free-space optical, and visible light communications. By comprehensively reviewing the literature on jamming in EBC, this survey paper aims to provide a valuable resource for researchers, engineers, and policymakers involved in the development and deployment of 6G networks. Understanding the nuances of jamming in different EBC technologies is essential for devising robust security mechanisms and ensuring the success of 6G communication systems in the face of emerging threats.
\end{abstract}


\section{Introduction}

\begin{figure}[t]
\centering
\includegraphics[width=3.3in]{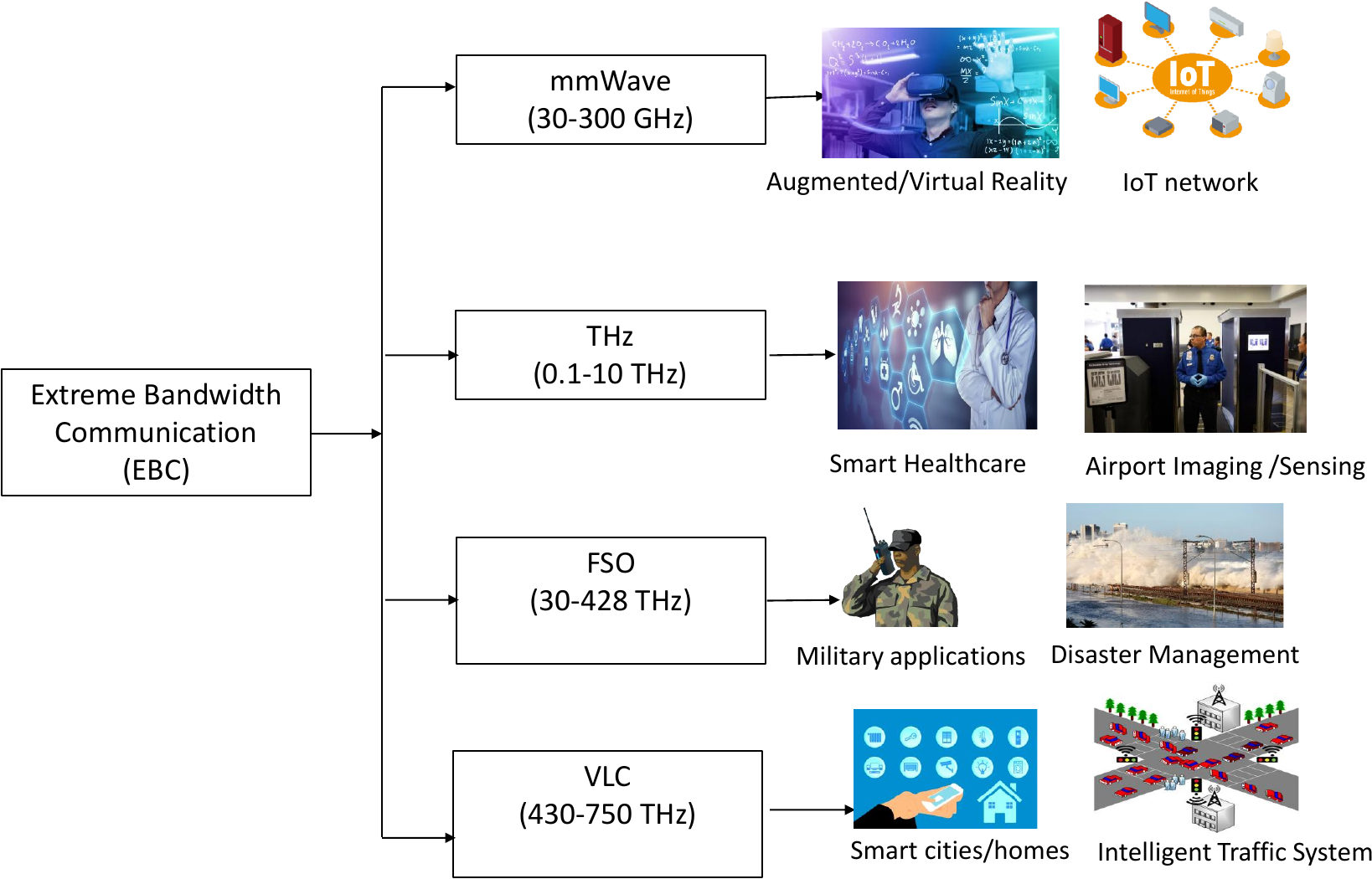}
\caption{ \small {Classification of Extreme Bandwidth Communication with potential 6G applications examples.}}
\label{f_1}
\end{figure}

\IEEEPARstart{R}{ecent} years have witnessed a remarkable upsurge in data-based internet usage, driven by the technological innovations of 5G/6G, machine learning, artificial intelligence, Internet of Things (IoT) devices, and more. Specially in the post-pandemic era, wireless communication has transitioned from being a luxury to becoming an essential everyday necessity. However, the future demands of wireless communications exceed the available resources within the current microwave ($\mu$W) radio-frequency (RF) spectrum in use. Because of this, researchers have proposed to pursue so-called "Extreme Bandwidth Communications" (EBC) which comprises of millimeter-wave (mmWave), terahertz (THz), and the optical bands (infrared and visible) \cite{6g_Mostafa, 6g_THz}. Based on the frequency band in use, EBC can be classified as mmWave communication, THz communication, free-space optical (FSO) communication, and visible light communication (VLC), see Fig. \ref{f_1}. 

MmWave communication is a kind of RF technology that uses frequency band of 30-300 GHz to achieve data rate of order of gigabit-per-second and offers huge bandwidth (BW) \cite{mmwave1}. MmWave communication has already been standardized in the 5G specifications, contributing to its current deployment in various industries for enhanced data transfer rates. With its proven utility in 5G, mmWave is poised to play a pivotal role in 6G, pushing the boundaries of connectivity with even higher data speeds and expanded applications in emerging technologies \cite{6g_Mostafa}. Further, moving to even higher frequency band than the mmWave, there is an emerging wireless communication technology called THz communication which operates in the THz frequency range of 0.1-10 THz. Compared to mmWave range, THz frequency range allows for significantly wider BW, and, therefore, the potential for much higher data rates. Similar to the mmWave, THz signals also have limited propagation range due to atmospheric absorption and scattering \cite{6g_THz}. 

Optical wireless communication (OWC) band (including infrared and visible) represents another spectrum capable of offering extensive BW resources \cite{Uysal2016, Majumdar2005,Agarwal207}. FSO and VLC are two distinct yet complementary technologies with potential roles in 6G networks. FSO communication uses infrared band of 30-428 THz and VLC uses the visible band of 430-750 THz. FSO is a line of sight (LoS) technology that transmits data using modulated beam of infrared light through free-space, making it suitable for scenarios demanding ultra-fast and secure data transfer. On the other hand, VLC utilizes visible light, often from light emitting diode (LED) sources, for communicating from one point to another, especially in the settings where there already exists lighting infrastructure (such as indoor and vehicular communication). Both FSO and VLC offer several advantages over the existing RF technologies such as unlimited unlicensed spectrum, ultra-high data-rate, less power requirement, and less electromagnetic interference. In the 6G landscape, where the emphasis lies on pushing the boundaries of connectivity, OWC's ability to provide high BW and low-latency communication is particularly valuable.

\textcolor{black}{It is clear that as we delve into the era of 6G and beyond where everyone and everything will be connected, EBC will serve as the driving force behind the innovations yet to be envisioned by offering unprecedented data rates and very low-latency connectivity \cite{6g_Mostafa}. Nevertheless, this hyper-connected wireless network will also introduce security issues such as jamming, eavesdropping, denial-of-service (DoS) attack, spoofing, message falsification attack, etc. Out of all these security threats, jamming and eavesdropping are two types of wireless attacks at the physical layer \cite{security_survey_hanzo}. Eavesdropping means covertly intercepting and extracting legitimate information from the channel; and jamming refers to the intentional disruption of the legitimate communication by causing interference \cite{survey-grover, security_survey_hanzo}. In the recent years, eavesdropping in EBC has been studied a lot, yet jamming still remains a less explored area \cite{eve-THz, eve-fso1, eve-fso2}. }
\begin{table}[h!]
\caption{List of Abbreviations}
\begin{tabular}{ll}
Abbreviation & Definition   
\vspace{0.2em}\\
AN           & Artificial Noise                           \\
AoA          & Angle of Arrival                           \\
AP           & Access Point                               \\
AT           & Atmospheric Turbulence                     \\
AWGN         & Additive White Gaussian Noise              \\
BER          & Bit Error Rate                             \\
BHT          & Binary Hypothesis Testing                  \\
BJM          & Blind Jamming Mitigation                   \\
BTS          & Base Transceiver Station                   \\
BW           & Bandwidth                                  \\
CST          & Carrier Sensing Time                       \\
CSK          & Color Shift Keying                         \\
DSSS         & Direct Sequence Spread Spectrum            \\
DoS          & Denial of Service                          \\
EBC          & Extreme Bandwidth Communication            \\
FHSS         & Frequency Hopping Spread Spectrum          \\
FoV          & Field of View                              \\
FSO          & Free-Space Optical                         \\
GHz          & Gigahertz                                  \\
GLRT         & Generalized Likelihood Ratio Test          \\
HAP          & High Altitude Platform                     \\
IoT          & Internet of Things                         \\
ITS          & Intelligent Transport System               \\
LDPC         & Low Density Parity Check                   \\
LED          & Light Emitting Diode                       \\
LoS          & Line of Sight                              \\
MCR          & Multi-Channel Ratio                        \\
MIMO         & Multiple-Input Multiple-Output             \\
MMSE         & Minimum Mean Square Error                  \\
mmWave       & Millimeter Wave                            \\
MRC          & Maximum Ratio Combining                    \\
$\mu$Wave       & Microwave                                  \\
OBUs         & Onboard Units                              \\
OFDM         & Orthogonal Frequency Division Multiplexing \\
OWC          & Optical Wireless Communication             \\
PD           & Photodiode                                 \\
PER          & Packet Error Rate                          \\
PHC          & Policy Hill Climbing                       \\
RBF          & Radial Basis Function                      \\
RIS          & Reconfigurable Intelligent Surface         \\
RF           & Radio Frequency                            \\
RSS          & Received Signal Strength                   \\
RS           & Reed Solomon                               \\
RSUs         & Roadside Units                             \\
RTS          & Request to Send                            \\
RZF          & Regularized Zero Forcing                   \\
Rx           & Receiver                                   \\
SOP          & Secrecy Outage Probability                 \\
THz          & Terahertz                                  \\
Tx           & Transmitter                                \\
UE           & User Equipment                             \\
UAV          & Unmanned Aerial Vehicle                    \\
V2X          & Vehicle-to-X                               \\
VANETS       & Vehicular Adhoc Networks                   \\
VLC          & Visible Light Communication                \\
WSN          & Wireless Sensor Networks                   \\
WLAN         & Wireless Local Area Network                          
\end{tabular}
\end{table}
\begin{table*}[t]
\caption{Comparing This Jamming Survey Paper to Previous Survey Papers }
\centering
\begin{tabular}{|l|l|l|l|l|l|l|}
\hline
Reference                     & Year & Considered Technology                                                                                                                               & \begin{tabular}[c]{@{}l@{}}Factors for   \\ Jamming \\ vulnerability\end{tabular} & \begin{tabular}[c]{@{}l@{}}Jamming   \\ Attacks\end{tabular} & \begin{tabular}[c]{@{}l@{}}Jamming  \\ Detection\end{tabular} & \begin{tabular}[c]{@{}l@{}}EBC Specific \\ Mitigation\end{tabular} \\ \hline
{[}14{]} Mpitziopoulos et. al & 2009 & \begin{tabular}[c]{@{}l@{}}Only for RF-based \\ wireless sensor networks\end{tabular}                                                               & Not Applicable (N/A)                                                                                & Yes                                                          & \begin{tabular}[c]{@{}l@{}}Only for\\  RF-WSN\end{tabular}    & No                                                                 \\ \hline
{[}15{]} Pelechrinis et. al   & 2009 & \begin{tabular}[c]{@{}l@{}}Not EBC-oriented study.\\ Have studied various \\ jamming models\end{tabular}                                            & N/A                                                                                & Yes                                                          & Yes                                                           & No                                                                 \\ \hline
{[}8{]} Grover et. al         & 2014 & \begin{tabular}[c]{@{}l@{}}Not EBC-oriented study.\\ Jamming in general \\ wireless \\ networks is studied\end{tabular}                             & N/A                                                                                & Yes                                                          & Yes                                                           & No                                                                 \\ \hline
{[}7{]} Zou et. al            & 2016 & \begin{tabular}[c]{@{}l@{}}Only for RF-based \\ Wireless Networks\end{tabular}                                                                      & N/A                                                                                & Yes                                                          & No                                                            & No                                                                 \\ \hline
{[}12{]} Vadlmani et. al      & 2016 & \begin{tabular}[c]{@{}l@{}}Only for RF-based \\ Wireless Networks\end{tabular}                                                                      & N/A                                                                                & Yes                                                          & Yes                                                           & No                                                                 \\ \hline
{[}17{]} Singh et. al         & 2020 & Only THz                                                                                                                                            & Yes                                                                               & No                                                           & No                                                            & Yes                                                                \\ \hline
{[}16{]} Vaishnavi et. al     & 2021 & \begin{tabular}[c]{@{}l@{}}Not cosidered EBC.\\    \\ General  physical layer \\ security for 6G with \\ conventional RF is discussed.\end{tabular} & No                                                                                & Limited                                                      & No                                                            & No                                                                 \\ \hline
{[}18{]} Paul et. al          & 2022 & Only FSO                                                                                                                                            & Yes                                                                               & Yes                                                          & No                                                            & \begin{tabular}[c]{@{}l@{}}Limited \\ discussion\end{tabular}      \\ \hline
{[}13{]} Pirayesh et. al      & 2022 & \begin{tabular}[c]{@{}l@{}}MmWave only as EBC \\ Primary focus is on RF-based \\ Wi-Fi, cellular, Zigbee,\\  Bluetooth, GPS, etc.\end{tabular}      & Not for EBC                                                                       & Yes                                                          & Yes                                                           & No                                                                 \\ \hline
\cite{vlc12:survey} Blinowski &  2019  & VLC  & Yes &  Yes   & No   & No  \\ \hline
Our paper                     & 2024 & MmWave, THz, FSO, VLC                                                                                                                               & Yes                                                                               & Yes                                                          & Yes                                                           & Yes                                                                \\ \hline
\end{tabular}
\end{table*}
\section{Literature Review} Many survey papers have been published focusing on the jamming threats of RF-based wireless systems \cite{survey-grover, security_survey_hanzo, jamming-survey-satish, survey-pirayesh, WSN-survey-jam, dos-jam-survey}. A. Mpitziopoulos et. al.  \cite{WSN-survey-jam} have investigated the significant challenge of jamming in wireless sensor networks (WSNs), with a specific focus on the vulnerabilities of communication protocols employed in WSN deployments. They have examined various types of jammers (constant, random, deceptive, reactive, etc.) posing threats to WSN networks. They have also discussed jamming detection techniques and several countermeasures against them (such as proactive, reactive, and mobile-agent based countermeasures). While the survey of \cite{WSN-survey-jam} is focused mainly on jamming in WSNs, K. Pelechrinis et. al. have studied some of the most harmful jamming attacks that can be launched at physical (PHY) and medium access control (MAC) layers of any wireless network in \cite{dos-jam-survey}. They have also discussed some jamming detection and prevention schemes at both PHY and MAC layer. The authors in \cite{survey-grover} have presented a comprehensive overview of jamming in wireless networks from attacker's point of view. They have explained different types of jammers and possible anti-jamming techniques by highlighting how the placement strategies of jammers affect their detection. They have also emphasized that it is important that the anti-jamming techniques are able to detect the jammer in an energy-efficient manner. The survey of \cite{security_survey_hanzo} presents a very detailed study about wireless security threats and their mitigation techniques at different protocol layers ranging from application to physical layer against malicious attacks. Further, it reviews various security protocols of wireless networks such as Bluetooth, wireless local area network (WLAN), commonly called Wi-Fi, WiMax, and long-term evolution (LTE) standards. This paper also discusses physical layer security techniques against eavesdropping, various jamming attacks and their counter-measures. In \cite{jamming-survey-satish, survey-pirayesh}, a detailed survey is presented covering the radio jamming attacks and potential defense strategies across diverse wireless networks. For example, the authors in \cite{jamming-survey-satish} have categorized different jamming attacks and defense strategies by considering the type of wireless network impacted, the viewpoint of the attacker or defender, the game type employed to model the jamming issue, and the methodology used for solutions. The considered wireless networks in this paper are WLAN, WSNs, and ad hoc networks. An in-depth comprehensive review of jamming attacks and anti-jamming strategies across various wireless networks, including WLANs, cellular networks, cognitive radio networks, ZigBee networks, Bluetooth networks, vehicular networks, LoRa networks, radio frequency identification networks, global positioning system (GPS) system, millimeter-wave, and learning-assisted wireless systems are presented in \cite{survey-pirayesh}. The authors have also performed a thorough examination of current anti-jamming techniques in wireless networks, encompassing power control, spectrum spreading, frequency hopping, multiple-input multiple-output (MIMO)-based jamming mitigation, and jamming-aware protocols. Additionally, a literature review on jamming attacks and countermeasures within emerging wireless technologies such as mmWave communications and learning-based wireless systems is also presented in \cite{survey-pirayesh}. Recently, a brief survey on physical layer security of 6G has been discussed in \cite{vaishnavi-jam} where they have explored some of the anti-jamming strategies. However, this survey is not very comprehensive and does not address the specific jamming problems of various EBC technologies.

It is worth noting that all of these aforementioned survey papers consider jamming only in the conventional RF based wireless communication systems. The study of jamming has been overlooked in most of the EBC technologies, often viewed as directional. Only a few individual survey papers in \cite{THz-boonbane} and \cite{FSO1-survey} provide discussions on jamming in THz and FSO communication, respectively. In \cite{THz-boonbane}, an in-depth discussion about the pros and cons of using THz for secure 6G communication is provided. They have also proposed various strategies to deal with jamming/eavesdropping threats at physical/link/network layers.  \cite{FSO1-survey} explores the features and technical constraints of FSO which make it vulnerable to jamming. The authors have pointed some of the potential anti-jamming techniques including spatial diversity and machine-learning based techniques. However, these works individually focus solely on THz and FSO communications and provide no insight into other EBC technologies. A notable absence exists in the literature with regards to a comprehensive survey paper specifically addressing the challenges of jamming in various EBC technologies for 6G applications. With the advent of hyper-connectivity in 6G applications, jamming is poised to become a considerable threat in these technologies \cite{6g_walid}. Acknowledging security and reliability as central key performance indicators not only strengthens 6G's technological foundation but also fosters lasting trust in the dynamic landscape of wireless communication. Therefore, in this paper we are motivated to discuss various jamming threats and potential anti-jamming techniques in EBC for various 6G applications.

The key contributions of this paper are outlined as follows. Section III offers an overview of jamming, including its definition, classifications, and potential strategies for mitigation. In Sections IV through VII, the paper delves into the issue of jamming within the context of 6G, utilizing EBC technologies. It discusses the factors that make these technologies susceptible to jamming, presents real-world examples of 6G applications that are at risk, and explores potential mitigation techniques for mmWave, THz, FSO, and VLC communications. The paper concludes with Section VIII, summarizing its contributions and suggesting directions for future research. 

\begin{figure}[!t]
\hspace{-3.5cm}
\includegraphics[width=5in]{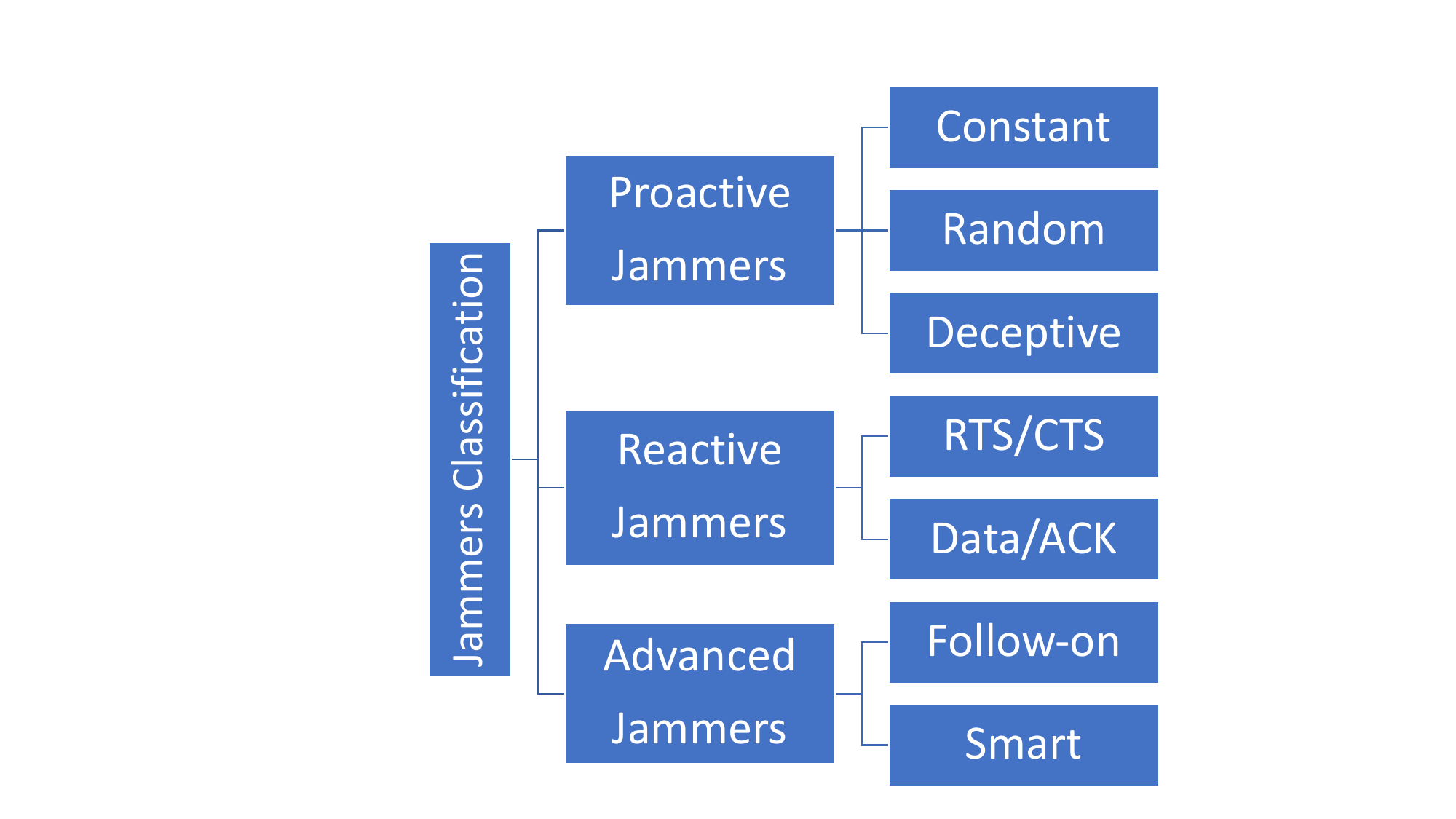}
\caption{\centering{Classification of jammers in EBC}}
\label{fig-jamtype}
\end{figure}
\section{Jamming: Definition and classification} \label{sec:jamtype}

\textcolor{black}{The history of jamming in wireless communication extends back to the early 20th century, notably during World War I and the Cold War. In these conflicts, jamming emerged as a strategic tool employed to intentionally interfere with enemy communication and radar systems, disrupting their ability to convey vital information  \cite{jamming-history-wiki}. As technology advanced, so did the sophistication of jamming techniques, impacting both military and civilian communication networks.}
\textcolor{black}{Jamming in EBC is defined as the deliberate disruption of wireless signals such as mmWave, THz, or optical signals, with the intention of preventing normal communication between wireless devices \cite{jam-def1, jam-def2}. Such interference can compromise the integrity of EBC, resulting in widespread service outages, data breaches, and compromised network reliability. Jamming is done by transmitting signal on the same frequency band used by the legitimate EBC system. Please note that jamming primarily occurs at the physical layer, and Fig. \ref{fig-jamtype} outlines the comprehensive classification of jammers predominantly disrupting the communication links of EBC. This classification is structured based on the technique employed for the jamming attack, with brief descriptions provided as follows:}
\begin{itemize}
    \item Proactive Jammer: Proactive jammers are designed to actively disrupt the communication links, regardless of whether legitimate transmissions have commenced or not. Proactive jammers can be further classified as Constant, random, and deceptive. 
    \begin{itemize}
        \item Constant Jammer: A constant jammer is a malicious device which persistently transmits a very high power random signal at specific frequencies in order to disrupt the legitimate wireless communication link. The random signals are continuously transmitted without following any carrier sense multiple access protocol of data transmission. This approach simplifies the implementation of constant jamming but is characterized by its energy inefficiency.
       \item Random Jammer: A random jammer transmits jamming signal at irregular intervals, alternating between "rest" and "jamming" periods. This method conserves some energy when compared to constant jamming. However, it results in relatively lower jamming effectiveness.
       \item Deceptive Jammer: Deceptive jammer is a kind of jammer which continuously sends meaningful regular information rather than sending random signals. By doing so, these jammers aim to deceive the receiver (Rx) into believing that a legitimate transmission is occurring. Similar to the constant jammers, they also have very low energy efficiency. However, the jamming effectiveness of deceptive jammer is higher than the constant jammer.
     \end{itemize}
    \item Reactive Jammer: Unlike constant jammers that transmit constantly, the reactive jammers constantly monitors the wireless channel and initiates transmission only after it senses some activity on the channel between legitimate transceivers. It remains inactive when there is no transmission taking place between the legitimate transceivers. Since they continuously monitor the channel and respond to the channel activity, reactive jammers are difficult to detect but they have low energy efficiency. Reactive jammers can jam the channel in two different ways: (a) Either it can transmit the jamming signal as soon as it senses a request to send (RTS) message from transmitter (Tx), thus distorting RTS message and stopping the Rx from sending a clear to send message. or (b) It can transmit the jamming signal towards the Rx to either corrupt the legitimate signal or corrupt the acknowledgement signal. In this case, the jammer sends the signal only after legitimate transmission has started. 
    \item Advanced Jammers: Advanced jammers are more adaptive and sophisticated in their jamming strategies than proactive and reactive jammers. Advanced jammers relevant to the EBC are:
    \begin{itemize}
        \item Follow-on Jammer:  Follow-on jammers can be detrimental to the EBC links which employ frequency-hopping/wavelength-hopping as an anti-jamming technique. When the Tx detects a jammer and try to switch to a different frequency band/wavelength, follow-on jammer has the capability to follow the Tx to the new frequency/wavelength.
        \item Smart Jammers: Smart jammers are quite energy-efficient and are capable of hindering communication with magnified jamming.  
    \end{itemize}
\end{itemize}

The jamming interference can jeopardize critical applications of 6G, such as autonomous vehicles, smart cities, tele-medicine, and real-time industrial automation, which rely heavily on seamless and high-speed connectivity. Given the exceptionally high frequencies and data rates expected at EBC in 6G applications, the consequences of jamming attacks are amplified. In the era of 6G where our everyday lives and business will rely on connectivity services, an attack on these services can cause havoc by posing a threat to data integrity within wireless communication systems. Therefore, for successful implementation of 6G in the coming years, it is mandatory to ensure security of the wireless system through jamming-mitigation.
\begin{table}[]
\caption{Summary of Potential Jamming-Mitigation Techniques}
\label{anti-jam}
\resizebox{\columnwidth}{!}{%
\begin{tabular}{|l|l|l|}
\hline
\begin{tabular}[c]{@{}l@{}}Jamming-Mitigation \\ Technique\end{tabular} & Overview                                                                                    & Effective Against                                                                \\ \hline
\begin{tabular}[c]{@{}l@{}}Regulated \\ Transmit Power\end{tabular}     & \begin{tabular}[c]{@{}l@{}}Optimizing \\ transmit power\end{tabular}                        & Reactive Jammer                                                                  \\ \hline
\begin{tabular}[c]{@{}l@{}}Spread-spectrum \\ techniques\end{tabular}   & FHSS, DSSS                                                                                  & \begin{tabular}[c]{@{}l@{}}Constant, Reactive, \\ Random, Deceptive\end{tabular} \\ \hline
Spatial Diversity                                                       & \begin{tabular}[c]{@{}l@{}}Use of multiple \\ Txs/Rxs\end{tabular}                          & \begin{tabular}[c]{@{}l@{}}Constant, Random,\\ Reactive\end{tabular}             \\ \hline
Jamming Filtering                                                       & ZF, MMSE                                                                                    & \begin{tabular}[c]{@{}l@{}}Deceptive, \\ partially for Reactive\end{tabular}     \\ \hline
\begin{tabular}[c]{@{}l@{}}Adaptive \\ Coding/Modulation\end{tabular}   & \begin{tabular}[c]{@{}l@{}}LDPC, RS, \\ strategic use of \\ modulation schemes\end{tabular} & \begin{tabular}[c]{@{}l@{}}Random, Deceptive\\ Reactive\end{tabular}             \\ \hline
AN generation                                                           & \begin{tabular}[c]{@{}l@{}}Use of \\ friendly jammer\end{tabular}                           & \begin{tabular}[c]{@{}l@{}}Deceptive,\\ Reactive\end{tabular}                    \\ \hline
RIS                                                                     & \begin{tabular}[c]{@{}l@{}}Optimizing \\ RIS Phase/amplitude\end{tabular}                   & \begin{tabular}[c]{@{}l@{}}Constant, Deceptive,\\ Random, Reactive\end{tabular}  \\ \hline
Game theory                                                             & \begin{tabular}[c]{@{}l@{}}Strategic game design\\ for jamming-mitigation\end{tabular}      & \begin{tabular}[c]{@{}l@{}}Smart, Reactive, \\ Deceptive\end{tabular}            \\ \hline
ML, AI                                                                  & \begin{tabular}[c]{@{}l@{}}Use of intelligence\\ for jamming-mitigation\end{tabular}        & \begin{tabular}[c]{@{}l@{}}Smart/Advanced,\\ Reactive, Deceptive\end{tabular}    \\ \hline
\end{tabular}%
}
\end{table}
\subsection {Potential Jamming-Mitigation Techniques} \label{jam_rm}
Jamming mitigation is a two step process: Jammer detection and jamming prevention/anti-jamming.
\begin{itemize}
    \item Jamming Detection: A jammer can be detected by identifying the abnormality in the received signal at the legitimate Rx using one of the following methods \cite{jamming-survey, security_survey_hanzo, mm-detection2} \footnote{Note that the aforementioned jamming-detection techniques are applicable to all EBC technologies.}:
    \begin{itemize}
        \item Received Signal Strength (RSS) Detector: A threshold based detector can be used to detect the presence of a constant, deceptive, random, and reactive jammer by measuring the RSS. Jammer can also be detected by Comparing the RSS of signals received during two consecutive transmissions. 
        \item Carrier Sensing Time (CST) Monitoring: Denial of service on physical layer because of jammer can occupy the channel and results into an unusual increase in CST. CST monitoring can detect constant, deceptive and random jammer. However, it can not be used to detect the presence of reactive and adaptive jammers.
        \item Packet Error Rate (PER): PER is defined as the ratio of unsuccessfully decoded data packets to the total number of received packets. Abnormally increased level of PER can detect the presence of constant, random, and reactive jammers; but can not expose a deceptive jammer.
        \item Noise Level Measurement: Jammer presence can be detected by differentiating between normal and abnormal ambient noise levels in a channel \cite{surfing}.
        \item Direction-based Detector: High angular resolution property of massive-MIMO wireless communication systems can be exploited to obtain accurate directional information of the Txs and the jammer. Using the directional information obtained during jammer detection, signal detection scheme can be designed to mitigate the jamming also \cite{mm-detect3}.
        \item Machine-learning Based Detector: High detection accuracy can be achieved by using machine learning based detection techniques. Classification problem can be formulated by using algorithms such as random forest, support vector machine, neural network, etc. with the input parameters of RSS, CST, bad packet ratio to determine the presence of jammer \cite{mm-detection2, 85-pirayesh}. Accuracy of designed detector can be assessed using parameters such as probability of detection, false alarm, and missed detection. 
    \end{itemize}

    \item Anti-Jamming: Following jamming prevention techniques can be considered for an EBC system \cite{survey-pirayesh},\cite[TABLE VI]{security-6g61}.
    \begin{itemize}
\item Regulated Transmit Power: Optimizing the transmit power can help in jamming-mitigation. For example, low transmit power of legitimate Tx can make it non-discoverable by the jammer. In case the jammer has been detected by the legitimate transmitter, legitimate signal to jamming power ratio can be improved by increasing the transmit power of the legitimate Tx, thus mitigating the impact of intended jammer \cite{jamming-survey-satish, THz-boonbane, WSN-survey-jam}. 
    \item Spread-Spectrum Techniques \cite{THz6, 4-THz6}: 
    \begin{itemize}
        \item Frequency Hopping Spread Spectrum (FHSS):  FHSS is a modulation technique where the Tx and Rx rapidly switch between different frequencies within a predefined spectrum thereby improving the jamming resistance of the system.
        \item Direct Sequence Spread Spectrum (DSSS): DSSS aids in jamming mitigation by spreading the original data across wider BW using a pseudo random spreading code.
        \item Hybrid FHSS/DSSS:  Hybrid FHSS/DSSS system can leverage the advantages of both frequency hopping and direct sequence spreading, offering a comprehensive solution for jamming mitigation across a wide range of potential threats.
        \item Differential DSSS:  In D-DSSS, instead of directly modulating the data with the spreading code, the changes or differences between successive bits of the spreading code are modulated onto the data. It enhances the robustness of DSSS against specific jamming scenarios by introducing differential modulation.
    \end{itemize}
    \item Spatial Diversity-based Techniques: MIMO systems can be leveraged for jamming-mitigation by employing various spatial processing techniques. For example, MIMO beamforming can be employed to direct the transmission or reception focus in specific spatial directions. Suitable precoder/combiner can also be used to cancel the jamming signal \cite{mmw2, Qyan}. Multi-channel ratio decoding can also be employed against constant and reactive jammers \cite{MCR}.
    \item Jamming Filtering at Rx: Jamming filtering can be performed at the Rx by following algorithms of regularized zero-forcing (RZF) and minimum mean square error (MMSE). RZF and MMSE can be used to estimate both legitimate and jamming channels. Based on the channel estimates, regularized linear receive filter can be designed \cite{jamming-survey}. In case of unavailability of channel information, blind jamming mitigation (BJM) can be performed by using the training sequence knowledge. BJM is effective when the number of receiving antennas at the Rx exceed the number of antennas at the jammer. 
    
    \item Error Correction Coding: Channel coding schemes are used to improve the link's reliability in presence of jammer. Low density parity check codes (LDPC) and Reed-Solomon (RS) codes have been proved to be quite effective against the jammers \cite{survey-pirayesh, 78-pirayesh}. Out of these two techniques, LDPC is more effective if the packet size is long and binary modulation scheme has been used.
    \item Dynamic Resource Allocation: Dynamic resource allocation involves dynamically assigning and managing resources, such as frequency bands, power levels, and time slots, to optimize performance in presence of jammers.
    \item Game-theoretic Methods: Game theory can be applied to jamming mitigation by modeling the interaction between the communication system and the jammer as a strategic game, where both parties aim to maximize their own outcomes. This approach enables the development of strategies that anticipate and counteract the jammer's actions, dynamically adjusting the system's parameters (such as frequency, power, and modulation) to maintain effective communication despite attempts at disruption.
    \item AN Generation: Once jammer is detected, a friendly jammer can be used to generate AN pointing towards the malicious jammmer.
    \item Use of Reconfigurable Intelligent Surface (RIS): By optimizing and adjusting RIS elements amplitude and phase, adaptive beamforming, selective signal enhancement, jamming signal nulling, and spatial diversity can be achieved.
    \item Modulation Schemes: Different modulation schemes can be strategically employed to mitigate jamming based on the sophistication and adaptability of the jammer. Adaptive modulation techniques, for instance, allow for dynamic switching between modulation schemes to counteract the jammer's efforts, enhancing communication resilience by adjusting to the jammer's intelligence level.
    \item Spatial Retreating Solution: Spatial retreat can be used as a defense mechanism to get away from the jammer. Mobile Tx can also be used so that sender can randomly move within Rx reception range (filed of view) to avoid the jammer \cite{Qyan}. An effective and continuous channel tracking and jamming detection is required to implement spatial retreat.

\end{itemize}
\end{itemize}
In the upcoming sections, we will delve into a comprehensive exploration of the jamming issue, examining potential techniques to mitigate jamming across diverse EBC technologies such as mmWave, THz, FSO, and VLC communications.

\begin{figure*}[!t]
\centering
\includegraphics[width=5in]{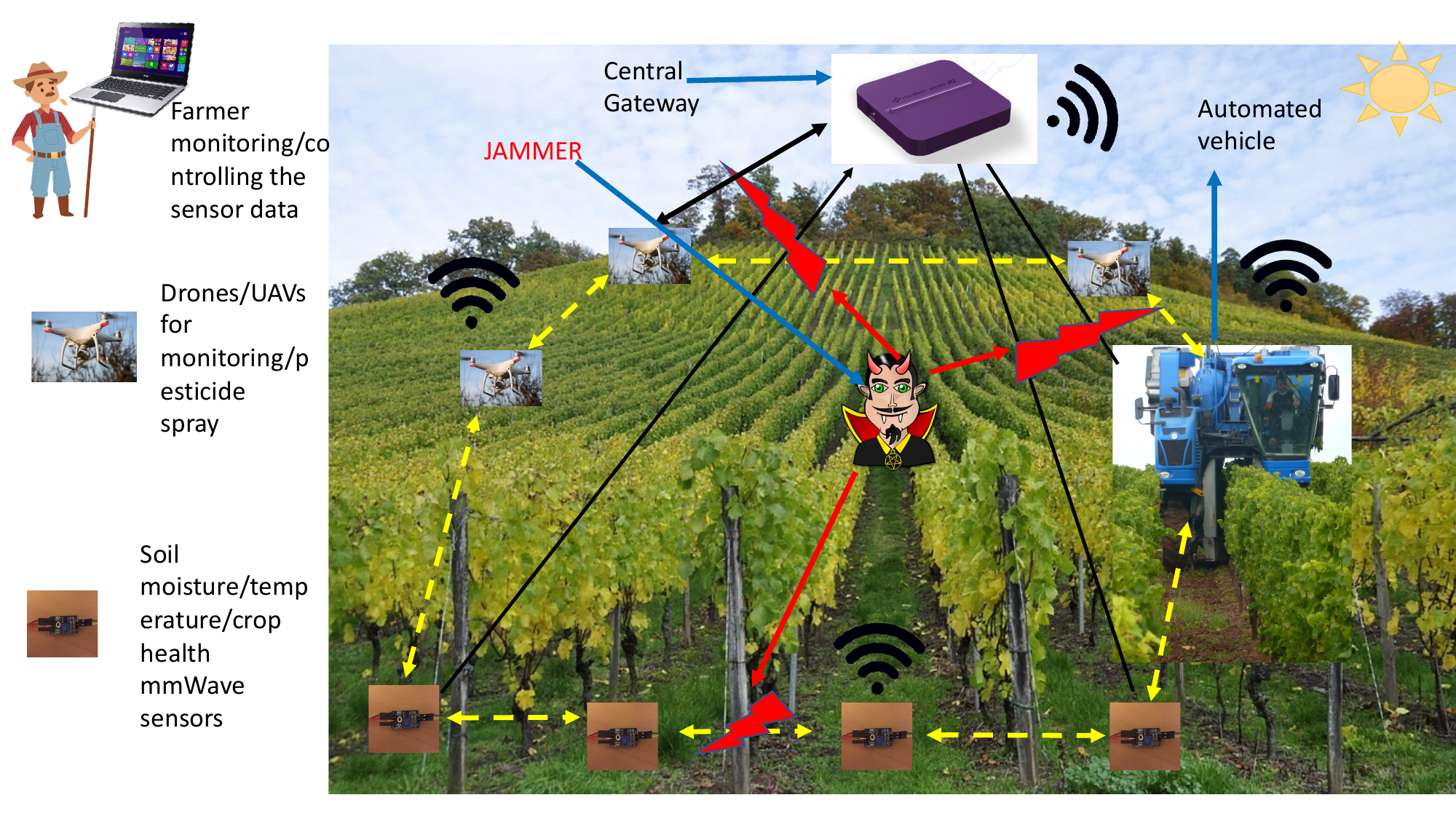}
\caption{\centering{Jamming in Smart Vineyard farming using mmWave Communication}}
\label{fig-mmwave}
\end{figure*}
\section{Jamming Threats in mmWave Communication}
\noindent Because of use of very high frequency, mmWave communication has very distinct characteristics from conventional $\mu$W RF systems. As opposed to traditional RF, mmWave signals have very short wavelength which allows them to transmit over a short range at a very high data rate. The shorter wavelength make them more prone to absorption by atmospheric gases, rain, and other environmental factors. Because of this, they become susceptible to attenuation and blockages by obstacle. To overcome the attenuation, mmWave usually uses highly directional antennas, therefore they are primarily considered as LoS technology. They employ directional beamforming which can be exploited by the jammers to disrupt the focused beams. The directional nature of mmWave transmission which is intended to enhance efficiency, can inadvertently render the system more susceptible to malicious interference.
\subsection{Factors Contributing to mmWave Communication's Susceptibility to Jamming}
MmWave communication, operating in the frequency range of 30 GHz to 300 GHz, possesses unique characteristics that can make it susceptible to jamming:

\begin{itemize}
    \item Limited Range: MmWave signals have shorter wavelengths, resulting in reduced propagation range. This characteristic makes the communication more susceptible to localized jamming attacks within the limited coverage area.
    \item Directional Transmission: MmWave systems often use directional transmission with narrow beams to overcome high free-space path loss. However, this makes them more susceptible to jamming, as interfering signals need to be aimed accurately to disrupt communication.
    \item Vulnerability to Obstacles: MmWave signals are sensitive to obstacles like buildings and foliage due to higher absorption and scattering. Jamming signals can exploit these vulnerabilities, using obstacles to attenuate or disrupt the mmWave communication link.
    \item Atmospheric Absorption: Certain atmospheric conditions, such as rain and oxygen absorption, can attenuate mmWave signals. Jamming attacks may leverage weather conditions to disrupt or degrade communication.
    \item Limited Penetration through Materials: MmWave signals have reduced penetration capabilities through walls and other materials compared to lower-frequency signals. This limitation makes it easier for jamming signals to obstruct mmWave communication by exploiting physical barriers.
    \item High BW and Data Rates: While the high BW of mmWave communication enables high data rates, it also makes the channel more susceptible to interference. Jamming attacks targeting specific frequency bands can significantly impact the overall communication quality.
\item Emerging Technology: MmWave communication is relatively new and rapidly evolving. The novelty of the technology may result in fewer standardized security measures, making it potentially more vulnerable to sophisticated jamming techniques.
\item Complex Beamforming Techniques: MmWave communication often relies on advanced beamforming techniques to overcome propagation challenges. Jamming attacks may attempt to exploit the complexity of these beamforming methods, disrupting the intended communication patterns.
\end{itemize}
\subsection{Practical Example of Jamming in mmWave communication} MmWave communication can be applied to high throughput and low latency communication in IoTs applications such as smart factory or smart agriculture. Following examples demonstrate the potential security challenges posed by jamming devices interfering with data communication in IoT based applications.
\begin{itemize}
    \item Smart Factory: In a smart factory IoT network, numerous IoT devices and sensors are strategically placed to monitor and oversee critical industrial processes. These devices continuously collect data on various parameters, such as temperature, pressure, humidity, and other real-time factors, ensuring product quality and safety. The network employs a single access point (AP) located centrally within the factory for the purpose of data collection and distribution. The AP facilitates communication with all the IoT devices. However, this setup introduces a security concern where one of the IoT devices may maliciously function as a jamming device, disrupting communication among the other devices. Such interference can lead to unreliable or lost data at the AP, potentially resulting in sub-optimal decisions concerning factory operations that could adversely affect product quality.
    \item Smart Agriculture: In the context of smart agriculture, the scenario involves multiple mmWave access points strategically placed throughout a farm to ensure comprehensive coverage, see Fig. \ref{fig-mmwave}. These access points enable communication with numerous mmWave sensors deployed across the agricultural land. The sensors are equipped with data collection and monitoring capabilities, collecting crucial information about factors like soil moisture, temperature, humidity, and crop health. These sensors transmit the collected data to the nearest mmWave Access Point (AP) for further processing and analysis. However, this agricultural setup also presents a security challenge, as a malicious actor with a jamming device can potentially target the AP. Such interference can result in data unreliability or loss at the AP, ultimately leading to sub-optimal farming decisions that could have detrimental consequences for crop yield and health \cite{smartagri}.
\end{itemize}
\begin{table*}[t]
\caption{Summary of jamming-mitigation in mmWave communications.}
\begin{tabular}{|l|l|l|l|}
\hline
EBC Technology           & Reference                                                                    & Technique                                                                                                                                    & Comment                                                                                                                                                                                                                                                                                                           \\ \hline
{MmWave} & \begin{tabular}[c]{@{}l@{}}{[}36{]} Zhang \emph{et. al},\\ 2023\end{tabular}        & \begin{tabular}[c]{@{}l@{}}Generalized-likelihood ratio test\\ (GLRT) for jamming detection\end{tabular}                                     & \begin{tabular}[c]{@{}l@{}}- One-step GLRT for detection under homogeneous noise \\ -two-step GLRT for detection under partially-homogeneous noise.\\ -Performs in scenarios with uncertain statistical noise info.\\ -Trade-off between detection performance \\ and computational complexity.\end{tabular} \\ \cline{2-4} 
                         & \begin{tabular}[c]{@{}l@{}}{[}24{]} Arjoune \emph{et. al},\\ 2020\end{tabular}      & \begin{tabular}[c]{@{}l@{}}ML-based detection.\end{tabular} & \begin{tabular}[c]{@{}l@{}}-Designed an ML-based classifier problem for jammer-detection.\\ -Studied random forest, support vector machine, \\ and neural network algorithm.\\ -Out of these 3, random-forest performs \\ the best with 97.5\% accuracy and low cost.\end{tabular}                                   \\ \cline{2-4} 
                         & \begin{tabular}[c]{@{}l@{}}{[}26{]} Bagherinejad \emph{et. al},\\ 2021\end{tabular} & \begin{tabular}[c]{@{}l@{}}Direction-based \\ detection/mitigation\end{tabular}                                                              & \begin{tabular}[c]{@{}l@{}}-Direction-based jamming detection\\and mitigation for massive MIMO.\\ -Identifies the presence and direction of jammer \\ by analyzing received pilot signal during training phase.\\ -Directional information of both legitimate user and jammer is used.\end{tabular}                    \\ \cline{2-4} 
                         & \begin{tabular}[c]{@{}l@{}}{[}31{]} Zhu \emph{et. al},\\ 2019\end{tabular}          & \begin{tabular}[c]{@{}l@{}}MIMO-based\\ hybrid beamforming\end{tabular}                                                                      & \begin{tabular}[c]{@{}l@{}}-Hybrid-beamforming to combat \\intended jamming in multi user (MU)-massive MIMO.\\ -legitimate signal can be recovered in presence of jammer.\\ -low computational complexity.\end{tabular}                                                                                                               \\ \cline{2-4} 
                         & \begin{tabular}[c]{@{}l@{}}{[}37{]} Xiao \emph{et. al},\\ 2016\end{tabular}                                                            & \begin{tabular}[c]{@{}l@{}}Reinforcement learning-driven\\ transmit power control\end{tabular}                                               & \begin{tabular}[c]{@{}l@{}}-Can work against smart jamming attacks.\\ -It is a learning-based optimal power allocation method. \\ -Studied for a downlink MU-massive-MIMO system.\end{tabular}                                                                                                                                                                                                             \\ \cline{2-4} 
                         & {[}38{]} Alsabet \emph{et. al}                                                      & RIS-based mitigation                                                                                                                         & \begin{tabular}[c]{@{}l@{}}-Multiple RISs are employed for jamming-mitigation.\\ -Secrecy rate is maximized by joint optimization\\ of all RISs phase shifts. \end{tabular}                                                                                                                                                                                                                                                       \\ \cline{2-4} 
                         & \begin{tabular}[c]{@{}l@{}}{[}39{]} Marti \emph{et. al},\\ 2021\end{tabular}        & \begin{tabular}[c]{@{}l@{}}Beam-slicing for\\ jamming mitigation\end{tabular}                                                                & \begin{tabular}[c]{@{}l@{}}--Practical technique designed for uplink MU-MIMO\\-Combats constant jammers.\\ -Beam-slicing plus a digital jamming mitigation\\ can provide increased jamming robustness.\end{tabular}                                                                                                                                                                                                                                                                                                                                                                                                                                                                                                                  \\ \cline{2-4} 
                         & \begin{tabular}[c]{@{}l@{}}{[}42{]} Hyppolite \emph{et. al},\\ 2023\end{tabular}    & \begin{tabular}[c]{@{}l@{}}Adaptive beam-management\\ to avoid the potential jammers\end{tabular}                                            & \begin{tabular}[c]{@{}l@{}}-Adaptive beam management solution to reduce jamming attacks.\\-Solution makes the system less discoverable by jammers.\\ -Reinforcement learning approach is used.\end{tabular}                                                                                                                                                                                                                                                \\ \hline
\end{tabular}
\end{table*}

\subsection{Jamming-Mitigation in mmWave Communication}
Authors in \cite{mm-detection1} have proposed method to detect jamming attack in a mmWave massive-MIMO system under composite noise uncertainty. Generalized likelihood ratio test (GLRT) is utilized to design a one-step GLRT and a two-step GLRT detector under homogeneous and partially homogeneous noise environment. It is shown through analysis that under homogeneous noise environment, a one-step GLRT jamming detector performs better than the two-step GLRT. Various jamming detection methods, leveraging machine learning, have been proposed using input parameters like bad packet ratio, packet delivery ratio, RSS, and clear channel assessment in \cite{mm-detection2}. Through the construction of a comprehensive dataset, the performance of random forest, support vector machine, and neural network algorithms has been evaluated, revealing the notably high accuracy of approximately $97.5\%$ is achieved by the random forest-based technique. Reference \cite{mm-detect3} introduces a direction-based approach for jamming detection and mitigation in mmWave massive MIMO systems. The proposed method identifies the presence and direction of the jammer by analyzing the received pilot signal during the training phase. Through utilizing the directional information of both legitimate users and jammers, a channel estimator is designed which is based on projecting received pilot signals onto the orthogonal complement of the jammer angular subspace. The channel estimate is then employed to design a combining vector for effective jamming mitigation.

Few works have explored MIMO-based techniques for jamming mitigation in a mmWave wireless communication system \cite{mmw2, xiao-mmwave, IRS-mitig-alsabet, G-marti, Ju-beamforming, Baron-SDR, Wang-AN}. In \cite{mmw2}, a hybrid beamforming technique is proposed to combat the effect of intended jamming in a mmWave-based multiuser massive MIMO system. Analog combiner of the Rxs are designed aiming to cancel the jamming effect by developing part of the analog matrix orthogonal to the jamming channel. The analog precoder matrix is calculated using the column iterative algorithm. The authors of \cite{xiao-mmwave} have introduced a reinforcement learning-driven power control algorithm designed for mmWave massive-MIMO systems. This algorithm utilizes fast policy hill climbing (PHC) algorithm to attain optimal transmit power in dynamic anti-jamming transmission scenarios, without requiring knowledge of the jamming and channel models. Simulation results demonstrate that the fast PHC-based power control strategy enhances overall utility of the mmWave massive-MIMO system against smart jammers. RIS is also being used to improve the physical layer security against jammers and eavesdroppers. For instance, multiple RISs are employed in a mmWave communication system in \cite{IRS-mitig-alsabet} where multiple links are considered between base station and the users to enhance the security against jammer and eavesdropper. They have proposed a method of maximizing the secrecy rate by jointly optimizing the phase shifts of all RISs and transmit beamforming. The authors of \cite{G-marti} have introduced a practical approach utilizing beam-slicing to effectively address strong jamming attacks on mmWave massive MU-MIMO systems equipped with low-resolution ADCs at the base station. Basically beam-slicing focuses jammer energy onto few analog to digital converters, so that some outputs are jamming-free. They have proposed beam-slicing based jamming-mitigation techniques called "soft-nulling of interferers with partitions in space" and "projeCtion onto ortHOgonal complement with partitions in space" which utilize the classical techniques of linear minimum mean square error equalization and projection onto the orthogonal subspace, respectively. Further, many works have explored using a friendly jammer to combat the impact of adversary jammer by introducing artificial noise (AN) into the channel. This strategic use of jammers serves the purpose of mitigating the adverse effects caused by intended jamming activities \cite{Ju-beamforming, Wang-AN}. A full-duplex jamming receiver is employed in  \cite{Ju-beamforming} to cancel the effect of self-interference and an eavesdropper in a multi-antenna based mmWave communication system. The work in \cite{Wang-AN} have studied the secrecy outage probability (SOP) of a mmWave system equipped with $N$ directional beamforming transmit antennas and a single omni-directional receiving antenna in presence of distributed eavesdroppers. It is demonstrated that by emitting AN in the side-lobes to cancel the eavesdroppers of side lobes areas, thus reducing the SOP and improving the system's performance.

\section{Jamming Threats in THz Communication}
\textcolor{black}{THz technology is at the forefront of emerging wireless communication, offering the potential for unprecedented data rates and capabilities. THz communication uses very high frequency range of 0.1-10 THz for data transmission. While the use of higher frequency offers benefits of higher data rate, they also come with very high path loss. To combat the path loss, usually high gain antennas are used which results into highly directional beams. Use of directional beam in THz communication contrasts with the conventional RF systems where the signals are typically broadcasted in a more widespread manner. The utilization of directional beams in THz communication instills a sense of security in the links, making it appear challenging for jamming attempts \cite{THz1}. By exploiting the narrow beams transmission in THz band, it might be possible to make wireless communication systems jamming-proof if the jammer resides outside the beam area. However, it is crucial to recognize that the issue of jamming remains a concern in THz spectrum, specially when the jammer is positioned within the beam sector or shares the same location as the Tx. Nevertheless, the relative novelty of THz technology also means that there is a paucity of comprehensive research and reported work focused on jamming within the THz frequency range. The authors in \cite{THz2} have reported  very interesting study addressing the security of THz links in presence of a jammer. They have shown, both theoretically and numerically that jamming is possible in THz communication if the jammer is able to target the sidelobes of the Rx by transmitting sufficient power (even though jammer is not in the exact LoS of the Tx and Rx). Further, It is shown in \cite{THz3} that, jammer can have quite detrimental effect even though it is not operating at the exactly same frequency as the Tx and Rx, as long as jammer's frequency is within the BW of the Tx. It shows that the use of wide transmission BW increases the chances of jamming and make it difficult to detect the jammer.
}

\textcolor{black}{ Recognizing the substantial emphasis on enhanced data security as a fundamental requirement in 6G communications, it is crucial to give due consideration to potential jamming threats within THz communication. A comprehensive examination of its implications is necessary to establish a robust communication system that can effectively counter such challenges.}

\subsection{Factors Contributing to THz Communication's Susceptibility to Jamming}
Following features of THz communication can be considered as the possible reasons that can make THz vulnerable to jamming:
\textcolor{black}{\begin{itemize}
\item 	Directional Beams: Use of highly directional beam, which makes it secure, can also be a reason for jamming if the jammer co-exists with the Tx and can accurately target the narrow beam.
\item Limited Range: Because the use of relatively short wavelength, THz communications have limited range capabilities. Because of this reason, it becomes easier for the jammers to place themselves within the effective communication range.
\item 	Narrow Frequency Range: THz communication typically functions within a limited frequency band, making it comparatively simpler for jammers to pinpoint.
\item 	Atmospheric Absorption: The signals in THz band are strongly absorbed by atmospheric gases like oxygen and water vapor. These absorption losses cause signal degradation and make THz communication more susceptible to jamming attacks.
\item	Emerging Technology: THz communication is still an emerging technology and development of effective anti-jamming techniques may lag the advancements of jamming technologies, leaving it more vulnerable to jamming threats. 
\end{itemize}
}
\begin{figure*}[!t]
\centering
\includegraphics[width=5in]{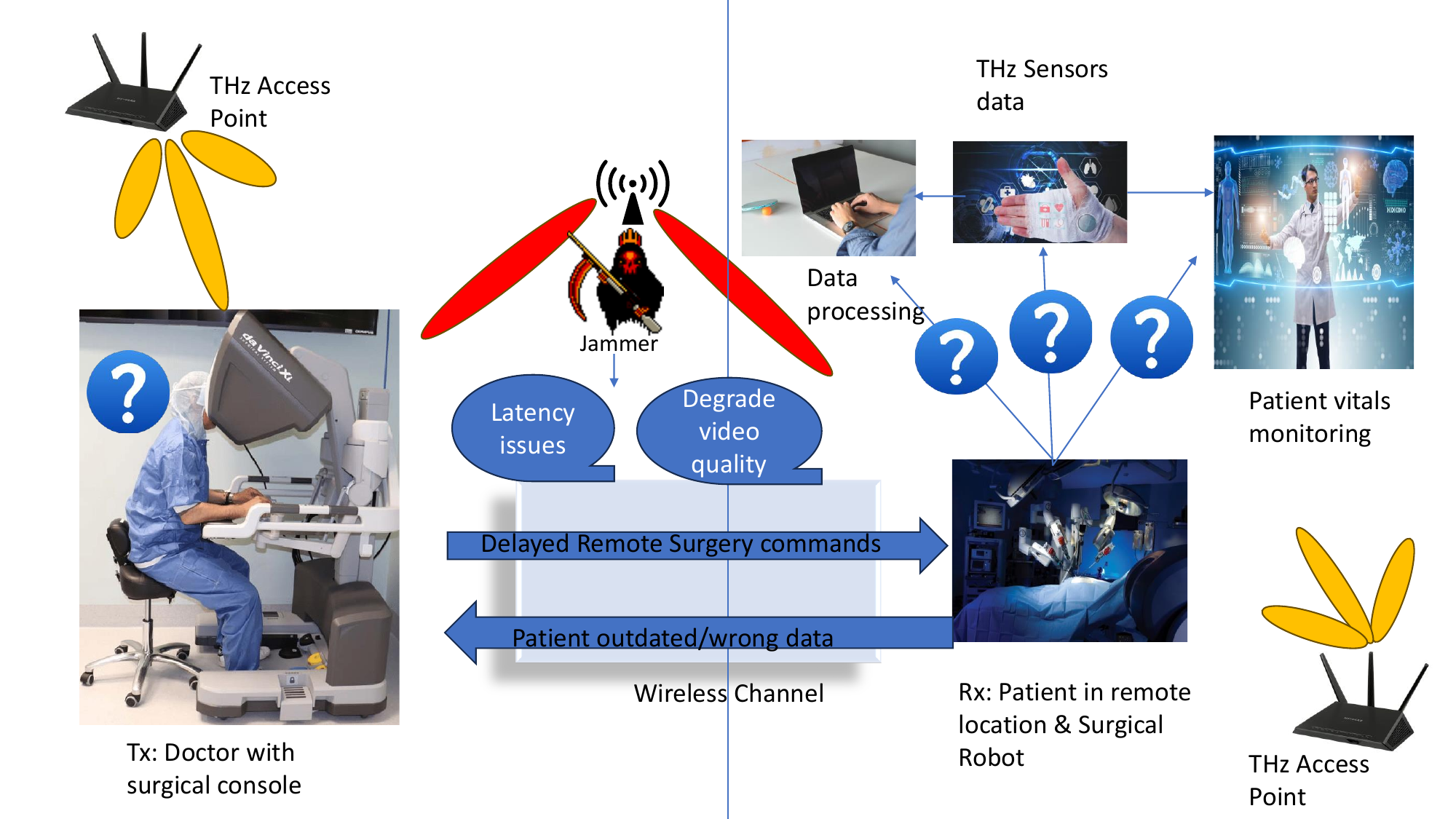}
\caption{\centering{Jamming in Remote/Robotic surgery using THz Communication}}
\label{fig-THz}
\end{figure*}
\subsection{Example of Practical Scenario/Application Susceptible to Jamming in THz Communication}
\textcolor{black}{\begin{itemize}
\item Remote and Precision Surgery: Jamming can indeed be a significant threat in remote precision surgery by causing communication disruption, especially when THz frequency communication is involved. In remote precision surgery, real-time communication between the surgeon's console and robotic surgical instruments is critical. THz communication may be used to transmit high-resolution video, control signals, and sensory data, as shown in fig. \ref{fig-THz}. A jamming attack can disrupt this communication, causing video lag, control delays, or complete signal loss in following manner.
\begin{itemize}
    \item Delayed Response: A jamming attack can introduce latency into the communication, leading to a delayed response between the surgeon's actions and the robotic instruments. Precision surgery requires millisecond-level accuracy, and delays can result in unintended movements or actions.
    \item Loss of Sensory Feedback: In remote surgery, sensory feedback is crucial for the surgeon to gauge tissue characteristics, apply the right amount of force, and detect abnormalities. Jamming can disrupt sensory feedback systems, making it difficult for the surgeon to make precise decisions.
    \item Potential for Harm: Jamming-induced delays or loss of control can lead to surgical errors, which can be harmful or even life-threatening to the patient. Precision surgeries often involve delicate procedures, and any disruption can have severe consequences.
    \item Data Privacy Concerns: Jamming attacks can also compromise the privacy and security of patient data. If the communication channels are jammed, sensitive patient information transmitted during surgery may be intercepted or disrupted.
6. Malicious Intent: Jamming attacks may be initiated with malicious intent, such as by disgruntled individuals, competitors, or those seeking to cause harm. These attacks can be intentional and targeted.
\end{itemize}
\item Some other practical THz applications which can be badly affected due to jamming attacks include medical/security imaging, and wireless backhaul. For instance, interruption in medical imaging can cause misdiagnosis thus compromising patient safety. Jamming attack on security imaging can  enable unauthorized individuals to bypass security checks unnoticed at airport, or public events. In case of wireless backhaul, jamming signals can overpower the legitimate THz signals, thus disrupting the communication between access network and core network. Prolonged jamming can also cause network outage, impacting the availability of wireless services and making the network unreliable.
\end{itemize}}
\begin{table*}[t]
\caption{Summary of JAMMING-MITIGATION in THz COMMUNICATIONS.}
\begin{tabular}{|l|l|l|l|}
\hline
EBC Technology & Reference & Technique                                                                                          & Comment                                                                                                                                                                                                                                                                                                                                                             \\ \hline
THz            & $\textrm{\cite{THz5} Gao \emph{et. al} }$  & Adaptive-modulation for jamming-mitigation                                                         & \begin{tabular}[c]{@{}l@{}}-Altitude dependent molecular absorption property \\ is exploited for unmanned-aerial vehicle (UAV) communication.\\ -Optimal carrier frequency and corresponding Tx power is \\ identified to maximize data rate.\\ -Works better in the downward channel.\end{tabular} \\ \hline
               & $\textrm{\cite{THz7} Gao \emph{et. al} }$   & \begin{tabular}[c]{@{}l@{}}Distance-adaptive absorption peak modulation\\ \\ (DA-APM)\end{tabular} & \begin{tabular}[c]{@{}l@{}}-Distance-dependent molecular absorption \\ property is exploited for jamming robustness\\ -Signals are modulated at molecular absorption peaks.\end{tabular}                                                          \\ \hline
               & {[}49{]} Mamaghani  & RIS-based technique for IoT network                                                                & \begin{tabular}[c]{@{}l@{}}-UAV-enabled RIS beamforming. \\ -Co-operative friendly jammer for \\AN-injection \\ -Signals are modulated \\at molecular absorption peaks\\ .\end{tabular}                                                                                                                                                                                                                                                                                                                                                                          \\ \hline
               & {[}50 {]} X. Pi & AN-injection using friendly jammer                                                                 &\begin{tabular}[c]{@{}l@{}}-Friendly jamming is employed \\ to interfere with adversary jammer. \\ -Makes the communication covert by degrading \\ detection performance of adversary jammer\end{tabular}                                                                                                                                                                                                                                                                                                                                                                     \\ \hline
               & {[}30{]} Nallappan  & \begin{tabular}[c]{@{}l@{}}Experimental spread spectrum-\\ FHSS using tunable laser\end{tabular}                                                        &\begin{tabular}[c]{@{}l@{}}-An experimental approach of FHSS  \\ -6 Gbps data rate over 1.75 m \\ with THz FHSS is demonstrated \end{tabular}                                                                                                                                                                                                                                                                                                                                                                      \\ \hline
               & $\textrm{\cite{4-THz6} Gao \emph{et. al} }$  &Spread-spectrum technique- Alternative FHSS     &\begin{tabular}[c]{@{}l@{}}-Optimal frequency is chosen at molecular absorption peaks\\ -For covertness, pulse waveform model\\ with polarization is developed. \\- Compared to normal FHSS, \\improvement in minimal covert distance by 60\% \end{tabular}                                                                                                                                                                                                                                                                                                                                                                       \\ \hline
\end{tabular}
\end{table*}
\subsection{Jamming-Mitigation in THz communication}
\textcolor{black}{It is noteworthy that THz path loss is typically modelled as sum of two components: free-space path loss and absorption losses due to water vapour and oxygen molecules. These molecular absorption losses within the THz band are frequency, distance, and altitude-dependent and the frequency-selectivity becomes more pronounced as distance increases. Consequently, it leads to the division of the spectrum into smaller segments, each with an width of the order of hundreds of GHz \cite{6g_THz, THz1, eve-THz}. The intrinsic properties of THz-based 6G systems inherently provide a certain degree of defense against jamming. These include an exceptionally narrow beamwidth with a frequency-dependent radiation pattern, distance-dependent and frequency-selective THz path loss, random small-scale mobility, and cell-free massive MIMO. Depending on the communication scenario, a combination of these features can be utilized to enhance the system's resilience against jamming \cite{THz1}. The works of \cite{THz5, THz7} have exploited these altitude and frequency-dependent molecular absorption properties of THz band to design jamming robust THz communication system. For instance, an adaptive modulation scheme is proposed in \cite{THz5} by exploiting the altitude dependent molecular absorption property of THz band to enhance the jamming robustness of an UAV communication system. The modulation scheme under consideration seeks to identify the optimal carrier frequency and corresponding transmission power with the goal of maximizing data rates while preserving covert operation. Simulation results indicate that the downward channel (Bob to Alice) provides greater covert capabilities compared to the upward channel. This distinction arises due to the lower density of water vapor observed at higher altitudes. Moreover, the work of \cite{THz7} explores a distance-adaptive absorption peak modulation where signals are dynamically modulated at the molecular absorption peaks to improve covertness in presence of an eavesdropper. Optimal carrier frequencies and allocated power are determined for rate maximization to reduce the eavesdropping distance, thus enhancing the covertness.}
\begin{figure*}[t!]
\centering
\includegraphics[width=5in]{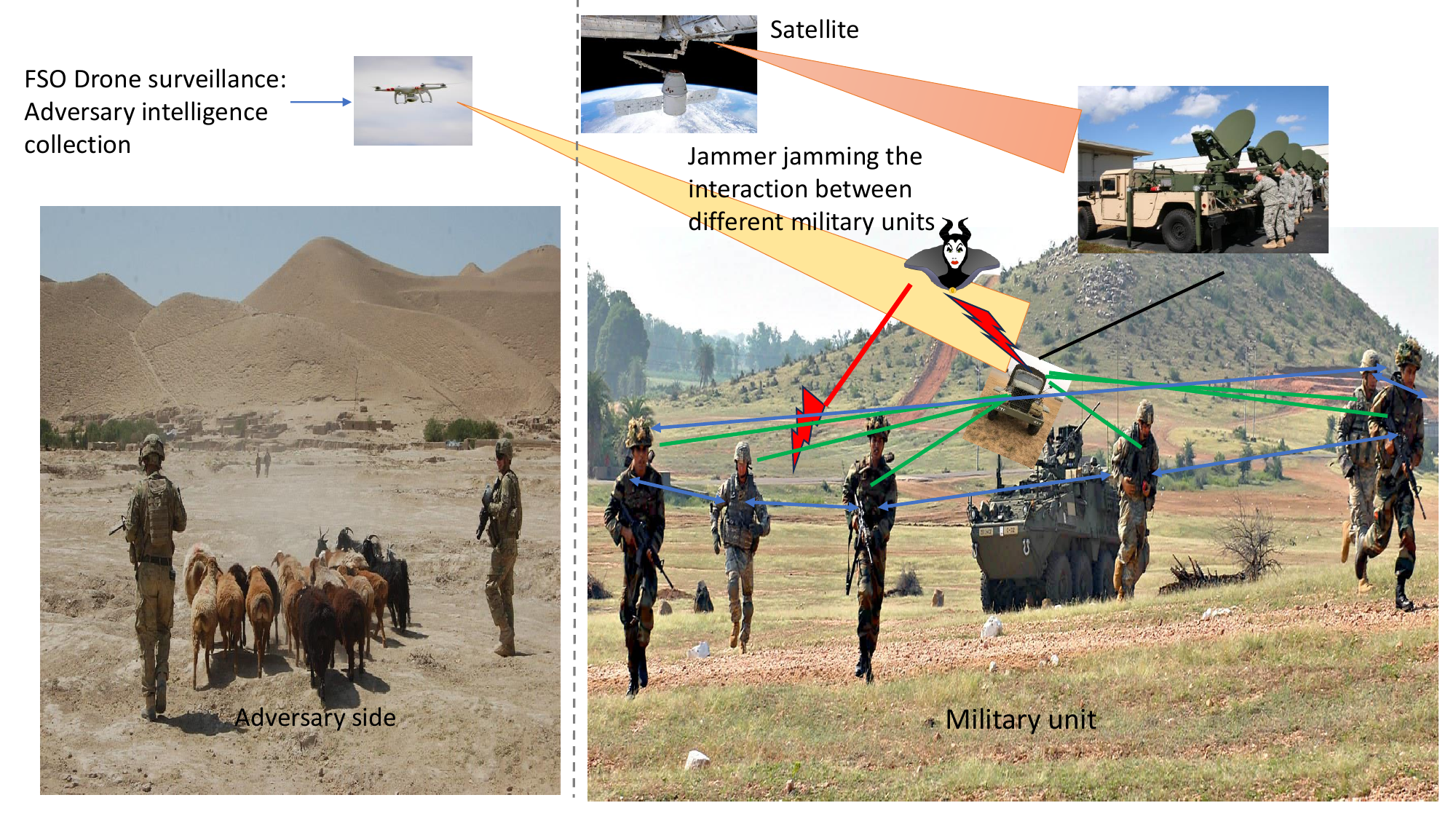}
\caption{\centering{Adversary jamming the secure FSO communication between the military units of an operation.}}
\label{fig4}
\end{figure*}
In the practical application of THz enabled IoT network with single AP and multiple user equipments (UEs), unscheduled UEs may behave as potential jammers. The authors of \cite{THz4} have proposed an energy-efficient secure covert IoT network communicating over THz band in presence of jammers. To ensure a secure link between AP and intended UE, a RIS-mounted UAV is used as a passive mobile relay. At the same time, a designated friendly jammer mounted on UAV is opportunistically used to transmit AN towards unscheduled UEs. The concept of friendly jamming has also been utilized in \cite{THz8} to enhance the channel security by combating the impact of eavesdropper/jammer in an UAV assisted two-hop THz network. They have assumed that two different jammers (J1 and J2) corresponding to both the hops separately. During first-hop transmission from Tx-UAV, UAV also sends AN towards J1 which lies between both of them. Similarly, in the second hop transmission from UAV-Rx, the UAV emits AN towards J2 to improve covertness of the socond-hop.

In the THz range, where numerous frequency channels are available due to the vast BW, frequency hopping can be implemented to provide security against jamming attacks \cite{THz6, 4-THz6}. In the study of reference \cite{4-THz6}, an alternative FHSS scheme has been investigated, wherein the optimal frequency is chosen strategically at the molecular absorption peak within the THz band. This innovative approach aims to bolster security measures against potential jammers. Moreover, the authors of\cite{THz6} have proposed an experimental solution of employing FHSS using tunable distributed feedback laser as a jamming mitigation technique for a photonics based THz communication.

Jamming in THz can be detected by using methods provided in Section III-A. However, to the best of our knowledge, while a few works exist on jamming-mitigation techniques, no one has explored jamming detection in literature yet.

\section{Jamming Threats in FSO Communication}
\textcolor{black}{Free-space optical (FSO) communication has gained a lot of popularity in the wireless communication applications as a complimentary technology to the conventional RF systems because of its features of unlimited BW, unlicensed spectrum, high throughput, low latency, low error rate, low power requirement, and negligible electromagnetic interference. FSO communication is widely recognized for its inherent security features, attributed to its LoS nature and utilization of narrow optical beams. FSO offers secure communication by transmitting data through free space using directed laser beams, making it challenging for eavesdroppers to intercept signals without being within the direct LoS. However, despite these security advantages, the performance of FSO can be impacted by jammers. Jamming in FSO involves intentional interference that disrupts the optical signals, compromising the reliability and security of the communication channel. Jammers can deploy various tactics, such as flooding the channel with intense light sources or introducing atmospheric disturbances, to disrupt the delicate transmission of optical signals. While FSO remains a secure communication option, the potential impact of jamming underscores the need for ongoing research and the implementation of countermeasures to fortify its resilience against intentional interference, especially in critical applications where secure communication is paramount.}

\subsection{Factors Contributing to FSO's Susceptibility to Jamming}
\textcolor{black}{Following factors can make FSO susceptible to the jamming attacks, which is inherently considered to offer secure communication.}

\begin{enumerate}
\item \textcolor{black}{Wide FoV Requirement: Literature establishes that FSO systems must employ spatial diversity techniques to maintain performance in the presence of atmospheric turbulence (AT) and pointing error (PE) \cite{PJ, access-richa}. A wide field of view (FoV) is essential to implement spatial diversity, allowing the Rx to capture beams from multiple Txs. The necessity of LoS imposes limitations on the communication range of FSO, leading to the deployment of UAVs or high altitude platforms (HAPs) to extend this range. For long-range FSO transmission using UAVs/HAPs, the direct LoS constraint is overcome, but the vibration and hovering of these platforms lead to angle of arrival (AoA) fluctuations in the optical Rx \cite{hap-richa}. Wide FoV Rxs can improve the reception probability of LoS beams in such scenarios. However, the downside is that a wide FoV creates additional space for jammers to position themselves between the Tx and Rx, enabling them to disrupt the link or initiate denial of service attacks \cite{isha-uavjam}. Mobility of UAVs/HAPs can also make FSO network more susceptible to jamming.}
    \item Unregulated BW of FSO: The absence of licensing requirements for transmitting signals on the infrared spectrum utilized by FSO communication introduces a heightened vulnerability to jamming. While this lack of regulatory constraints facilitates ease of deployment and operational flexibility, it simultaneously opens avenues for potential vulnerabilities. Unlicensed spectrum facilitates unrestricted access to the communication channel, allowing potential adversaries or unauthorized users to exploit the unregulated BW, posing a significant risk of intentional interference and jamming, as there are no legal barriers preventing malicious disruptions within this spectral space.
    \item Confined Range of Operating Wavelengths for Transmission: FSO communication operates within narrow wavelength ranges, primarily 780-850 nm and 1520-1600 nm \cite{fsona}. The latter range is favored in practical applications due to its ability to transmit more power, cover longer distances at higher data rates, and fall within the eye-safe spectrum (in contrast to the former range). This preference for a specific wavelength range could potentially make it easier for a jammer to identify and target the wavelength, posing a challenge to the security of FSO links.
\item Fixed Transmission Paths and Readily Identifiable:
FSO systems are prone to jamming due to their conspicuous positioning at elevated locations, typically installed on heights like building rooftops or towers, making them easily detectable and targetable. Moreover, FSO systems often have fixed transmission paths between those easily identifiable points. Jammers can anticipate and target these predictable paths, making it easier to disrupt communication between known locations.
\item Power Constraint: FSO systems may operate with limited power due to regulatory constraints, eye safety reasons or practical considerations. This power limitation can make the communication link more susceptible to intentional jamming efforts.
\end{enumerate}
\subsection{Practical Scenarios/applications Susceptible to Jamming in FSO}
Secure Military Operations: Consider an example of military operation where a special forces unit is deployed for a covert operation in a remote and hostile environment \cite{FSO-military}. For a secure and covert communication, FSO communication is being used as infrared beams are invisible to naked eyes, they can be rapidly deployed and also need low power for transmission \cite{fsona}. The unit requires a secure and high-BW communication network to exchange real-time data, such as video feeds from surveillance cameras, sensor data, and command instructions, among its members. However, the adversary somehow become aware of use of FSO communication and disrupts the special forces unit's links by deploying a skilled operative equipped with a sophisticated jamming device, see Fig. \ref{fig4}. Targeting the FSO system's wavelength, the jammer strategically interferes with optical signals, causing frequent disruptions, delays, and intermittent loss of communication among team members. The adversary's tactics compromise the unit's coordination, real-time data exchange, and response capabilities. In response, the special forces must swiftly adapt by implementing countermeasures to mitigate jamming effects and ensure the mission's success. The scenario underscores the vulnerability of FSO communication to intentional interference in military contexts and emphasizes the need for effective countermeasures to maintain operational effectiveness. 
\subsection{Jamming Mitigation in FSO} Out of all the EBC technologies, FSO is considered to provide the most secure communication link. Because of this reason, very little research has been done exploring the jamming-mitigation techniques for FSO communications. As mentioned in Section III-A, jamming mitigation is a two step process consisting of jamming detection and jamming prevention/anti-jamming.

 Encoding of transmit signals using cyclically shifted optical codeword can be used to detect a jammer which is injecting erroneous data into the multiple-layer UAV-FSO channel. When two UAVs communicate, they avoid using the codeword that identifies the sender UAV, making the communication more secure. The Rx decodes the data based on set of optical decoders configured based on cyclically shifted code replicas for bit 0 and 1. In the event of wrong decoding, presence of jammer is detected. Once the presence of jammer is detected, Rx UAV can mitigate the jamming attack by changing its spatial and temporal locations \cite {fsodet1}. Multi-channel ratio (MCR) decoding can also be employed at Rx to detect the presence of a constant jammer in an FSO-MIMO system using UAVs. In \cite{fsodet2}, the authors have considered an UAV network based on FSO communication with spatial diversity. It is considered that a constant jammer with single Tx is continuously injecting noisy signal into the network to jam single or multiple FSO links. Employing MCR-based detection, the Rx can identify the presence of the jammer and successfully recover the transmitted signals, even when subjected to a jamming attack. This detection method relies on analyzing the channel coefficients ratio associated with the jammer.

 Spatial diversity based techniques can be explored by employing multiple transmit apertures to combat the impact of a jammer. It has been shown through bit error rate (BER) and outage probability studies that employing multiple apertures at the transmitting end can help to mitigate the detrimental impact of a random jammer which lies within the Rx's FoV\cite{pratiti-MIMO}-\cite{pratiti-PJ}. The authors of \cite{isha-TAS} have coupled spatial diversity with transmit antenna selection method for jamming mitigation in FSO.

 To mitigate the effects of jammer and improve the secrecy performance of an FSO system, a friendly jammer can be used to direct AN towards the harmful jammer \cite{FSO-AN-isha, FSO-AN1}. Although the works of \cite{FSO-AN-isha, FSO-AN1} have used the friendly jammer to combat the impact of eavesdropper, nevertheless, similar concept can be applied to mitigate the jamming effects of FSO if the jammer/jammer's AoA is detected. 

 A buffer-aided relaying can be used to counter the impact of jamming in a co-operative FSO network by employing max-link selection protocol. Buffer-aided relaying in a co-operative FSO shows better jamming-resistant as compared to non buffer-aided co-operative FSO system in presence of jammer \cite{Buffer}. 
 As outlined in Section III-A, the use of RIS in FSO systems can effectively counteract jamming effects. For instance, the research presented in \cite{prakriti-irs-jam} investigates a cooperative FSO system with a dual-hop setup involving UAVs. In this scenario, an RIS is mounted on a second UAV positioned between the source and the relay node, aiming to achieve spatial diversity. The introduction of RIS facilitates spatial diversity by establishing multiple propagation paths between the transmitter (Tx) and receiver (Rx), thereby aiding in the mitigation of jamming effects.
 However, devising strategy to reduce the probability of jamming attacks is also a good mitigation strategy. As shown in \cite{fso-djord-arizona, fso-djord2}, it is practically possible to achieve a covert communication link in an outdoor FSO in presence of strong AT and jammer. A successful covert transmission of binary phase shift keying (BPSK) modulated, LDPC coded 10 Gbps signal, which is imposed on ASE noise source output signal over 1.5 Km has been demonstrated in \cite{fso-djord-arizona}. Adaptive optics (AO) has been used to mitigate the AT effects. Broadband incoherent light sources are used in \cite{fso-djord2} to achieve covertness while transmitting BPSK modulated signal in an outdoor FSO in presence of background solar radiations. 
\begin{table*}[t]
\centering
\caption{Summary of jamming-mitigation in optical wireless communications.}
\begin{tabular}{ | m{0.12\textwidth} | m{0.15\textwidth} | m{0.3\textwidth} | m{0.3\textwidth} | }
\hline
EBC Technology      & Reference         & Technique            & Comment                                   \\ \hline
 & $\textrm{\cite{pratiti-MIMO, pratiti-PJ} Paul \emph{et. al} }$     &  Spatial diversity                                 & Better BER and outage performance achieved using multiple Txs    \\
       \cline{2-4} 
&  $\textrm{\cite{isha-TAS} Chauhan \emph{et. al} }$     & Spatial diversity with transmit antenna selection                                  & Transmit aperture with maximum channel gain is selected for data transfer     \\
\cline{2-4} 
FSO &  $\textrm{\cite{Buffer} Shukla \emph{et. al} }$     & Buffer-aided relaying                                  & Buffer-aided FSO outperforms non-buffer aided in presence of jammer      \\
\cline{2-4} 
&  $\textrm{\cite{prakriti-irs-jam} Saxena \emph{et. al} }$     & RIS-based mitigation                                & RIS is employed to achieve spatial diversity and combat jamming     \\
\cline{2-4} 
&  $\textrm{\begin{tabular}[c]{@{}l@{}}\cite{fso-djord-arizona} Nafria\\ \cite{ fso-djord2} Djordjevic,\end{tabular} }$     & LDPC coding and incoherent light source                                 & BPSK modulation with LDPC coding is used for covert communication     \\
\hline
 &       &                                   & $\bullet$ Reduced interference       \\
      &                                 & $\bullet$ Hyperchaotic baseband frequency hopping                                        & $\bullet$ Robustness against noise    \\
      &  $\textrm{\cite{vlc7} Al-Moliki \emph{et. al}}$                               & $\bullet$ Optical-OFDM based on chaotic codes     & $\bullet$ Bandwidth inefficiency  \\
VLC      &                                 & $\bullet$ Secret key encryption                                                     & $\bullet$ Latency due to hopping \\
      &                                 &                                                                                     & $\bullet$ Increased complexity of design \\
 \cline{2-4} 
       &  $\textrm{\cite{vlc11} Chang \emph{et. al}}$       & \multirow{2}{0.3\textwidth }{Integrated arrayed waveguide grating router and optical switch for implementing wavelength hopping } & $\bullet$ Increased complexity of design      \\
             &                                 &                                                                                     & $\bullet$ Inherent nature of wide wavelength spectrum of light signals  \\

\hline
\end{tabular}
\end{table*}

\section{Jamming Threats in VLC}
VLC networks originally utilize light sources such as light emitting diodes (LEDs) which can emit the white light in the visible light wavelength around the range from 350 nm to 780 nm for illustration as well as communication \cite{vlc1, vlc2}. VLC has unique advantages over other conventional RF systems. First, VLC is pursuing dual purpose, illumination and communication supporting larger BW in unlicensed spectrum band. Second, VLC is robust to the security issue, i.e., ``\emph{What you see is what you get} (WYSIWYG)''. Third, LED-based VLC consumes lower power. Last but not least, the light-based emission does not harm human beings. Nevertheless, the VLC is vulnerable to the interference emitted by other ambient light sources. Since the photodiode (PD)-based Rx responds to the whole VLC wavelength range, any the artificial light sources or natural sunlight exposing to the VLC Rx can degrade the signal quality. There is a challenge to the vulnerability of jamming attack due to the broadcasting and superimposing nature of intensity signals in VLC systems. Furthermore, smart lighting system which can be remotely controlled can easily manipulated as an intended or unintended jammer that infringes on the network accessibility of VLC systems deployed with adjacent LEDs. Generally, VLC systems are targeted by classical type of jammers mentioned in Section \ref{sec:jamtype}, e.g., constant, reactive, deceptive, random, and periodic jamming attacks. The constant jamming attack, also known as illumination attack in the VLC systems, is most commonly considered on VLC jamming~\cite{vlc4,vlc5,vlc6,vlc7} in which LED, fluorescent or incandescent lamp illuminates the light with constant average intensity. Attacker may hijack a part of VLC infrastructure as a rogue Tx who emits jamming light~\cite{vlc4}. This kind of rogue Tx can send deceptive jamming signals, also referred to as impersonation attack~\cite{vlc5} to decoy the legitimate VLC Rx. Cleverly, the jammer exploits impersonation jamming attack for idle VLC channel, while constant jamming attack for busy VLC channel. Noise jamming that emits the additive white Gaussian noise (AWGN) signals has been investigated in~\cite{vlc7}, even though it is not exactly AWGN random jammer due to the non-negative intensity constraint of VLC signals. 

Advanced jammer can exploit the systematic information on VLC specifications to distract the VLC link more effectively from the perspective of energy efficiency. Jamming the preamble and header in IEEE 802.15.7 VLC frame format~\cite{vlc8} might be an energy-efficient adversary signaling activity. If the jammers can obtain a prior knowledge about the VLC systems and are disguising as legal users, they can attack the time synchronization or frame detection by using false preamble attack, preamble nulling attack or preamble warping attack. VLC-specific non-negative real-valued optical orthogonal frequency division multiplexing (OFDM) schemes have been widely studied to improve the spectral efficiency~\cite{vlc9:ofdm}, while the jamming attack on frequency synchronization is efficient to thwart sampling frequency offset compensation. Selective jamming in frequency domain is efficient to distract signals in RF systems, while the one in wavelength domain is efficient in VLC systems. This is because the response characteristic in wavelength highly depending on optical transceiver specification is affecting the system performance. Therefore, the VLC systems based on color shift keying (CSK) are more susceptible to wavelength jamming which can emit the malicious signal spanning whole wavelength range in VLC. Correspondingly, it is imperative to investigate the specific factors which make VLC systems vulnerable to adversary jamming signals.

\begin{figure*}[t!]
\centering
\includegraphics[width=5in]{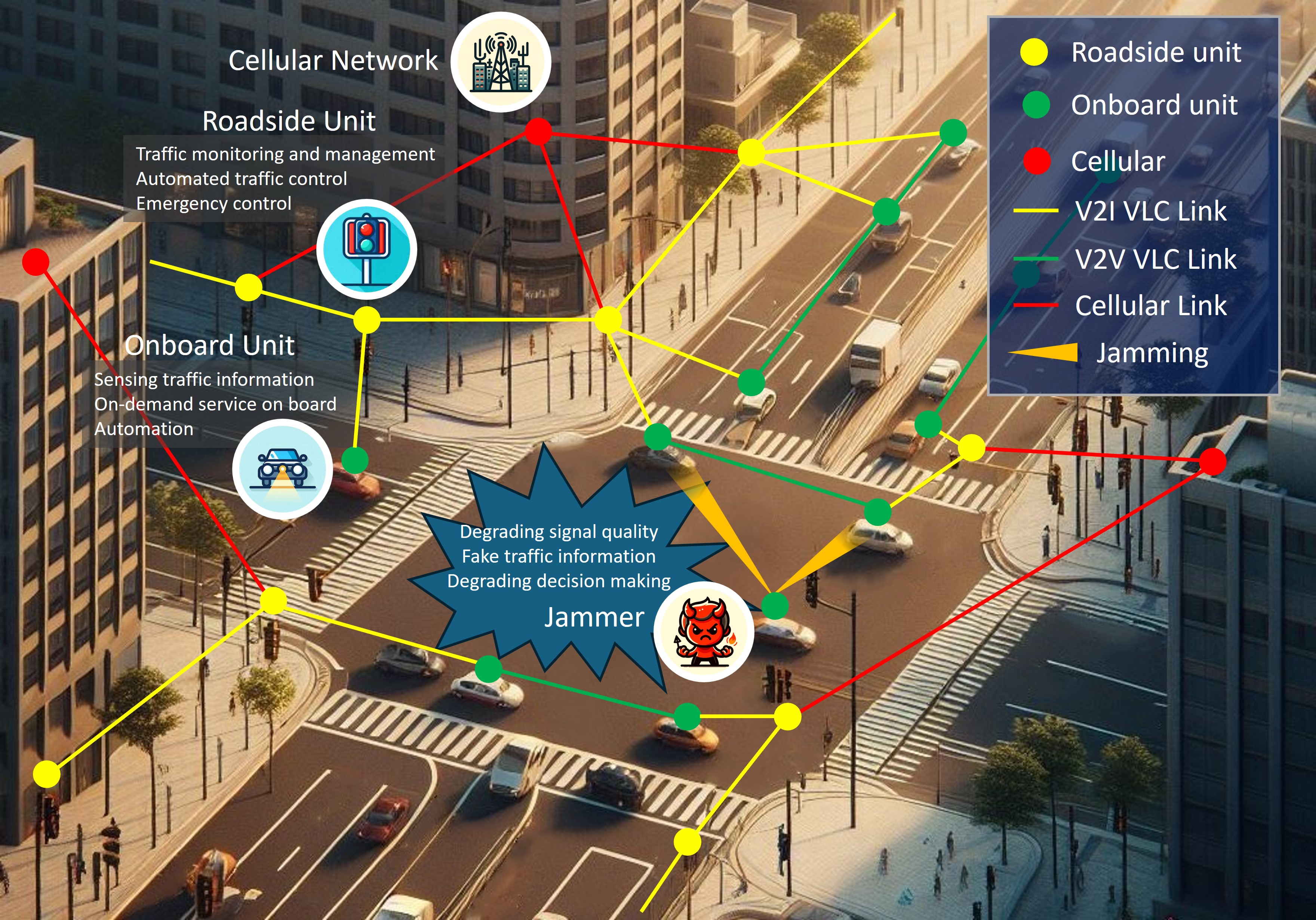}
\caption{\centering{Adversary jamming in VLC-based network for intelligent transportation systems.}}
\label{fig5}
\end{figure*}
\subsection{Factors Contributing to VLC’s Susceptibility to Jamming}
The specific features of VLC systems significantly considered as the factors to make VLC potentially susceptible to jamming are as following:
\begin{itemize}
    \item Spectrum of optic components: Most commonly used white LEDs and the Rx component such as silicon PIN or avalanche photo-detector have very wide wavelength response around 400-1000 nm which includes the whole visible light and near-infrared~\cite{vlc10:led}. Any lighting sources which is intended or not intended to emit ambient light such as sunlight, fluorescent light, LED light from computer/television monitors can become a devastating interference source leading to significant degradation of performance loss. This means that any types of lighting sources can be an adversary jammer in the full VLC wavelength region.

    \item Property of optic components: VLC Rxs feature on wide field-of-view (FoV) single PD or a structured narrow FoV multi-PD spanning wide incident angle. It is beneficial to collect and detect the desired signal when the Rx is located off the beam center, while it is potentially increasing the chance to interfere by harmful jamming signals. Therefore it is critical to adapt the Rx aperture to balance the trade-off between the signal reception and jamming mitigation.

   \item Network topology: The VLC network in indoor scenario is often spatially closed, in other word, the VLC users in the space are solely connected to one AP which is vulnerable to a jamming and cannot help rerouting data traffic in the jammed regions. In outdoor scenarios such as VLC-based intelligent transport system (ITS) networks, the users which are called onboard units (OBUs) are of mobility and continue to move one to another AP which is referred to as roadside units (RSUs). Moreover, the OBUs are continuously changing the network topology varying the positions and number of connected users. The smart jammers estimate the control channel and transmission policy of ITS networks, correspondingly send the disguised or replayed signals, induce the ITS networks to utilize the specific transmission policy, and attack it.  


\end{itemize}

\subsection{Practical Scenario/Applications Susceptible to Jamming in VLC}
ITS systems are intelligently operating on the base of vehicular ad-hoc networks (VANETs) supporting vehicle-to-X (V2X)  communications with a variety of sensing data, traffic and road assistance information in order to improve the efficiency of transport systems and services such as smart traffic control and information transfer, autonomous and unmanned driving, OBU applications, etc~\cite{vlc3, vanet1}. VLC systems through roadside lighting infrastructure and vehicle's head/tail lamps could be an important network components in ITS systems to exchange the vehicle and traffic information. The malicious VLC node endeavors to simply send the jamming signals to block the transmission between networked OBUs or between networked OBU and RSU AP. A smart jammer, equipped with intelligent radio devices, can monitor the ongoing VANET communication and assess the underlying policies. This not only enables the jammer to flexibly control jamming frequencies and signal strengths but also allows it to influence VANET to adopt a specific communication strategy, enabling targeted attacks. For example, the jammer can smartly send disguised or fake information to manipulate or sabotage the ITS system. With false decision making with fake information, selfish OBU may induce the artificial traffic flow which is advantageous to itself. Ultimately, it can cause a catastrophic traffic disruption that paralyzes either traffic in either local or wide area. As illustrated in these examples, jamming devices disrupting data communication present potential security challenges.

%

\subsection{Jamming Mitigation in VLC}
Generally, a jammer's presence can be identified by detecting anomalies in the received signal at the legitimate Rx by employing one of aforementioned methods in Section~\ref{jam_rm}. Specifically in VLC, binary hypothesis testing (BHT) has been employed at the legitimate Rx to detect the impersonation attack where Eve pretends to deceptively act as Alice~\cite{vlc5}. Direct current (DC) channel gain is exploited as fingerprint of the transmit LEDs (Alice) for the feature-based authentication. Cooperative jamming detection algorithm has been proposed in~\cite{vlc6}, where a classifier distinguishes whether a jamming attack is causing the degradation in bit error rate (BER) of cooperative LEDs. Jamming attack was detected by extracting the features and patterns of BER deterioration. 

On other front, when the jamming attack is detected, the VLC systems appropriately employ the jamming mitigation techniques at either Tx or Rx sides. General mitigation techniques in Section~\ref{jam_rm} can be leveraged in VLC systems. In~\cite{vlc7}, the authors proposed a hyperchaotic baseband frequency hopping based on optical orthogonal frequency division multiplexing (Hyperchaos-BB-FH-OOFDM) scheme to strengthen the availability and confidentiality in the face of constant, deceptive, and random jamming attacks. The proposed method is embedding chaotic time/frequency scrambling, subcarrier phase scrambling, and chaotic spreading codes. Wavelength hopping scheme has been implemented by employing arrayed waveguide grating router and optical switch~\cite{vlc11} and the random hopping signals generated by pseudo random noise generator algorithm were encrypted well in the wavelength domain and robust to the jamming.

\section{Conclusion and Future Research Direction}
To conclude, this article thoroughly explores the risks of jamming in 6G using EBC technologies such as mmWave, THz, FSO, and VLC communications. The article begins by defining jamming, categorizing it, and introducing common anti-jamming techniques. The discussion delves into factors influencing susceptibility to jamming in various EBC technologies, with practical examples like smart farming and remote surgery to highlight potential threats in 6G applications. Additionally, the article summarizes existing anti-jamming efforts for each technology, identifying research gaps and challenges in the field. Ultimately, the work underscores the need for advanced research to ensure the development of secure and reliable jamming-resistant 6G technologies.

Based on the insights provided, the field of jamming threats in emerging communication technologies for 6G applications presents multiple avenues for future research. Here are some specific directions proposed:
  \begin{itemize}
      \item Spread-Spectrum Techniques for mmWave Communications: While significant work has been done on detecting and mitigating jamming in mmWave communications through spatial diversity, beamforming, RIS, and ML techniques, spread-spectrum methods remain unexplored. Investigating these techniques could provide novel insights into jamming mitigation strategies.
      \item Advanced Mitigation Techniques: The exploration of jamming mitigation techniques such as filtering, advanced coding schemes, AN injection, and Game-theoretic approaches has been identified as an area needing further research across different communication technologies.
      \item Jamming Detection in THz Communications: There is a noticeable gap in research on jamming detection within THz communication. Developing methods for reliable jamming detection is crucial for maintaining communication integrity and demands immediate attention. Techniques from DSSS, ML, Game theory, coding schemes, and spatial diversity could be adapted and explored for THz communication systems.
      \item OWC Technologies: OWC, including FSO and VLC, have seen minimal research in the context of jamming. For FSO, exploring new jamming detection methods such as noise level measurements, directional detectors, and ML-based detectors could provide fresh perspectives. Additionally, the effectiveness of encoding/decoding techniques for jamming detection warrants further examination.
      \item Spatial Diversity in FSO: Limited studies have explored spatial diversity for jamming mitigation in FSO. There's potential to use RIS not just for achieving spatial diversity but also for optimizing phase and amplitude adjustments to enhance secrecy rates and reduce error rates in the presence of jammers.
      \item Comprehensive Techniques in OWC: In Optical Wireless Communications, including both FSO and VLC, there's a significant opportunity to explore advanced and effective techniques for jamming mitigation. These could include spread-spectrum methods, advanced coding schemes, AN injection, beamforming, and jamming filtering. For instance, FHSS can be employed in OWC by using multiple lasers/LEDs for generation of different carriers.
      \item VLC Communication Exploration: For VLC communication, future research could focus on jamming detection methods like RSS, PER analysis, noise level measurement, and ML-based approaches. In terms of mitigation, investigating strategies such as spatial diversity, regulated transmission power, dimming control, RIS, and spatial retreat could offer new solutions to enhance security and reliability.

  \end{itemize}
  Addressing these research directions will not only fill the current gaps in knowledge but also significantly contribute to the development of more secure, reliable, and jamming-resistant communication technologies for the 6G era.

\bibliography{ref}
\bibliographystyle{IEEEtran}

\end{document}